\documentclass[apjl]{emulateapj}
\usepackage{comment}
\usepackage{ifthen}
\usepackage{amsmath}
\usepackage{amsfonts}
\usepackage{amssymb}
\usepackage{multirow}
\usepackage{wasysym}
\usepackage{subfigure}
\usepackage{epsfig}
\usepackage{float}

\newcommand{\luna}{{\tt LUNA}}
\newcommand{\multi}{{\sc MultiNest}}
\newcommand{\cofiam}{{\tt CoFiAM}}
\newcommand{\kepler}{{\it Kepler}}

\newboolean{emulateapj}
\setboolean{emulateapj}{true}

\newboolean{color}
\setboolean{color}{true}

\shortauthors{Kipping et al.}
\shorttitle{Kepler-90\lowercase{g}.01 is a False Positive}
\ifthenelse{\boolean{emulateapj}}{
    
}{
    
}

\begin{document}

%% Titlepage
\title {
The Possible Moon of Kepler-90\lowercase{g} is a False Positive
}

%% Authors
\author{
	{\bf D.~M.~Kipping\altaffilmark{1,2},
             X.~Huang\altaffilmark{3},
             D.~Nesvorn\'y\altaffilmark{4},\\
             G.~Torres\altaffilmark{1},
             L.~A.~Buchhave\altaffilmark{1,5},
             G.~\'A.~Bakos\altaffilmark{3,6,7},
             A.~R.~Schmitt\altaffilmark{8}
	}
}

\altaffiltext{1}{Harvard-Smithsonian Center for Astrophysics,
		Cambridge, MA 02138, USA; email: dkipping@cfa.harvard.edu}

\altaffiltext{2}{Cfa Menzel Fellow}

\altaffiltext{3}{Dept. of Astrophysical Sciences, Princeton University,
		Princeton, NJ 05844, USA}

\altaffiltext{4}{Dept. of Space Studies, Southwest Research Institute, 
1050 Walnut St., Suite 300, Boulder, CO 80302, USA}

\altaffiltext{5}{Centre for Star and Planet Formation, Natural History Museum of 
                 Denmark, University of Copenhagen, DK-1350 Copenhagen, Denmark}

\altaffiltext{6}{Alfred P. Sloan Fellow}

\altaffiltext{7}{Packard Fellow}

\altaffiltext{8}{Citizen Science}

%% EOF authors

% #####################################################################
%% abstract
\begin{abstract}

The discovery of an exomoon would provide deep insights into planet formation 
and the habitability of planetary systems, with transiting examples being 
particularly sought after. Of the hundreds of \kepler\ planets now discovered,
the seven-planet system Kepler-90 is unusual for exhibiting an unidentified
transit-like signal in close proximity to one of the transits of the long-period
gas-giant Kepler-90g, as noted by \citet{cabrera:2014}. As part of the ``Hunt
for Exomoons with Kepler'' (HEK) project, we investigate this possible exomoon
signal and find it passes all conventional photometric, dynamical and centroid 
diagnostic tests. However, pixel-level light curves indicate that the moon-like
signal occurs on nearly all of the target's pixels, which we confirm using a 
novel way of examining pixel-level data which we dub the ``transit centroid''. 
This test reveals that the possible exomoon to Kepler-90g is likely a false 
positive, perhaps due to a cosmic ray induced Sudden Pixel Sensitivity Dropout
(SPSD). This work highlights the extreme care required for seeking non-periodic 
low-amplitude transit signals, such as exomoons.

\end{abstract}

% #####################################################################
%% keywords
\keywords{
	techniques: photometric --- stars: individual (Kepler-90) --- 
        planets and satellites: general --- planetary systems
}

%% EOF keywords
%% EOF titlepage

% #####################################################################
%% Introduction
\section{INTRODUCTION}
\label{sec:intro}

The discovery of a confirmed extrasolar moon remains an outstanding challenge
to modern astronomy. Despite nearly 2000 exoplanets having now been discovered
(www.exoplanet.eu; \citealt{schneider:2011}), with some smaller than Mercury
\citep{barclay:2013}, the confirmed existence of a satellite to any of these 
planets eludes us.

Both gravitational microlensing and transit photometry are predicted to be 
sensitive to ``large'' exomoons (masses $\gtrsim0.1$\,$M_{\oplus}$) and yet
even a plausible candidate system remains elusive. As an example of the 
sensitivity from microlensing, \citet{bennett:2014} recently identified a system
(MOA-2011-BLG-262Lb) which is either composed of a Neptune orbiting a late 
M-dwarf or a terrestrial moon ($\sim0.5$\,$M_{\oplus}$) orbiting a 
super-Jupiter, with the moon scenario statistically disfavored\footnote{We
define a ``candidate'' exomoon to be a case where the moon hypothesis is 
favored but it has not been unambiguously demonstrated to be uniquely explained 
by said hypothesis.}. Using transits, our project (the ``Hunt for Exomoons with
Kepler''; HEK) has published exomoon constraints for 17 transiting 
planets with six yielding sub-Earth mass constraints \citep{hek:2013,
kepler22:2013,hek:2014}.

Recently, \citet{cabrera:2014} and \citet{schmitt:2014} announced the discovery 
of the first example of a seven-planet transiting system, Kepler-90 (KOI-351).
Notably, \citet{cabrera:2014} discuss the possibility of the 210.6\,d period
planet, Kepler-90g, hosting a large exomoon, largely driven by a moon-like 
transit signal neighboring the third transit of this gas giant. Although 
\citet{cabrera:2014} remained cautious and did not claim to have found the first 
exomoon candidate, the visually apparent moon-like signal remains a mystery and 
arguably the most plausible evidence for a transiting exomoon in the published 
literature. For this reason, as part of the on-going HEK project, we here 
present an investigation of the possible exomoon of Kepler-90g, dubbed 
Kepler-90g.01 in what follows.

\section{PHOTOMETRIC ANALYSIS}
\label{sec:photometry}

\subsection{Background}
\label{sub:background}

\begin{figure*}
\begin{center}
\includegraphics[width=17.0 cm]{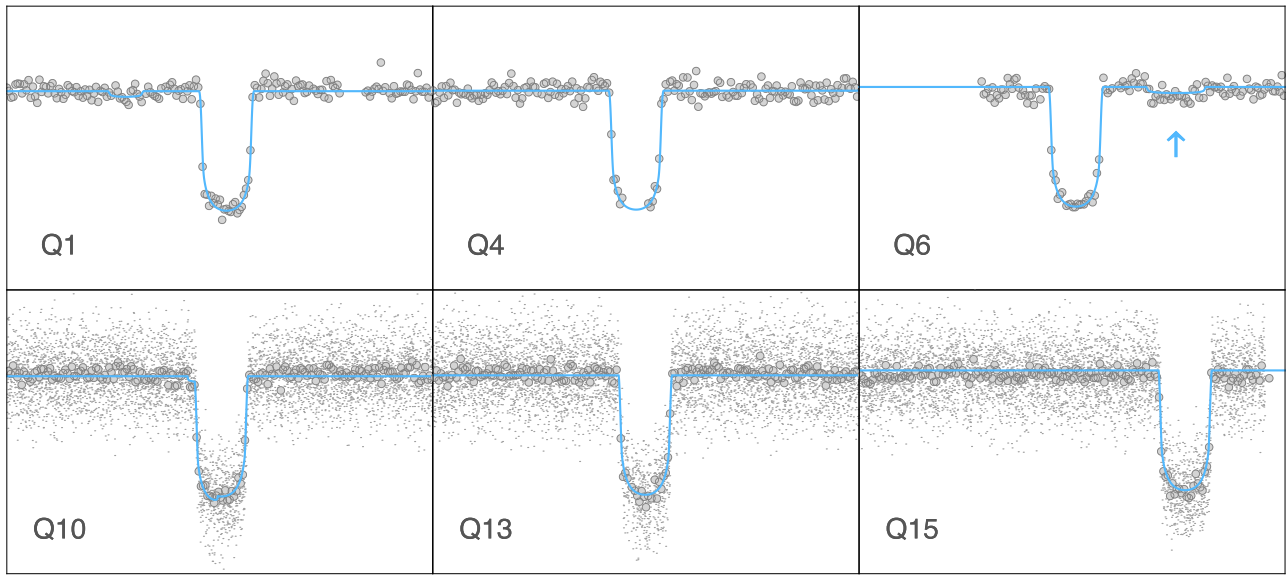}
\caption{\emph{
Photodynamical light curve fit (solid line) of the six transits of Kepler-90g,
including an exomoon and freely fitted transit times. Filled circles represent 
the LC data, and dots represent the SC data (where available). Temporal 
baseline ($x$-axis) of each caption is 96\,hours, centered on the times of 
transit minimum for a linear ephemeris. Relative flux ($y$-axis) plotted from 
$0.993$ to $1.003$. The Kepler photometry of Kepler-90g is well-explained by an 
exomoon plus additional TTVs.
}} 
\label{fig:lightcurve}
\end{center}
\end{figure*}

We begin our investigation by exploring whether the full set of photometric
observations of Kepler-90g can be explained by a planet-moon configuration,
which was not attempted by \citet{cabrera:2014}. There are two reasons why one 
might suspect this is not possible. Firstly, there is apparently only one 
high-significance piece of evidence for Kepler-90g.01, namely the companion 
transit to Kepler-90g in Q6 (quarter 6), as shown in Fig.~\ref{fig:lightcurve}. 
This raises the question as to whether it is dynamically possible for 
Kepler-90g.01 to avoid imparting a second transit elsewhere for the other five 
transits of Kepler-90g. Secondly, the transit of Kepler-90g.01 is 
$\sim21.5$\,hours later than that of Kepler-90g, implying a large planet-moon 
separation. Indeed, this is the principal reason why \citet{cabrera:2014} do not 
consider the moon hypothesis likely, arguing the moon would be on the edge of 
Hill stability.

\subsection{Data Treatment}
\label{sub:detrending}

The seven-planet system, Kepler-90, was observed continuously by the \kepler\
spacecraft from Q1 to Q17, except for Q8, Q12 and Q16 where the
target fell off the CCDs. We use all of the available data for this
target, including the short-cadence observations where available (from Q9 
onwards). The Pre-search Data Conditioning (PDC) time series \citep{smith:2012}
was downloaded and detrended after first excluding all transits, outliers and 
manually removing discontinuities (e.g. ramps), in the same way described in our 
previous papers (e.g. see \citealt{hek:2014}). Long-term detrending for the 
transit photometry of Kepler-90g was handled by the \cofiam\ algorithm 
\citep{hek:2013}, protecting timescales less than or equal to the full duration 
of Kepler-90g (12.593\,hours; Table 3 of \citealt{cabrera:2014}).

Kepler-90g exhibits a $\sim$1\,day transit timing variation (TTV) for its
sixth and final transit, which is far too large to be due to an exomoon alone
\citep{kipping:2009a}. For this reason, we model the six transits of Kepler-90g
using the photodynamical planet-moon code \luna\ \citep{luna:2011} plus six
freely fitted transit times, which are likely due to planet-planet interactions. 
By fitting the transit times freely, any TTVs due to an exomoon are absorbed by 
these parameters and thus would be degenerate with the moon's mass. This is 
particularly salient since TTVs due to an exomoon should dominate a mass 
determination over transit duration variations (TDVs; \citealt{kipping:2009a}), 
due to the visually obvious large separation between the transits of Kepler-90g 
and Kepler-90g.01, implying a large planet-moon separation. For these reasons,
the mass of the exomoon is simply set to zero in the photodynamical model (but
the moon radius is still freely fitted). As with our previous works, we regress
our model to the detrended photometry using the \multi\ package 
\citep{feroz:2008,feroz:2009}, with our priors following those of 
\citet{hek:2014}.

\subsection{Results: Formal Significance}
\label{sub:significance}

Using the Bayesian evidence computed by \multi, we find a planet+moon+TTVs
model is favored over a planet+TTVs model with a formal significance of 
$5.2$\,$\sigma$ ($\Delta\ln\mathcal{Z} = 15.3\pm0.2$). Furthermore, we find
that the six transits of Kepler-90g are well-explained by this model, with
Kepler-90g.01 imparting two dominant signals on the light curve, the originally
identified signal during Q6 and a second event preceding the first transit of
Kepler-90g (Q1), as shown in Fig.~\ref{fig:lightcurve}. 

The Q6 transit of Kepler-90g.01 is supported by a change in $\chi^2$ of 
$\Delta\chi^2=49.0$ over 26 data points. The Q1 transit is far less convincing, 
imposing a modest improvement of $\Delta\chi^2=11.1$ over 15 data points, during
a period of increased light curve scatter. We also note a very minor 
exomoon feature during the Q10 transit of negligible significance. We therefore 
conclude that the case for Kepler-90g.01 rests solely on the reality of the
Q6 moon-like transit.

\subsection{Results: Plausibility of Parameters}
\label{sub:plausibility}

The parameters for the planet-moon system, assuming 
$R_{\star}=(1.2\pm0.1)$\,$R_{\odot}$ \citep{cabrera:2014}, indicate a 
$(7.96\pm0.65)$\,$R_{\oplus}$ planet orbited by $(1.88\pm0.21)$\,$R_{\oplus}$
exomoon at semi-major axis of $(a_{SP}/R_P)=69.4_{-6.4}^{+7.3}$ planetary radii. 
We find short-period moons ($\lesssim10$\,days) are prohibited, but otherwise 
the moon's period posterior distribution is broad, leading to a wide range of 
possible planet-densities via the technique described in \citet{weighing:2010} 
of $\rho_P<21.4$\,g\,cm$^{-3}$ to 95\% confidence. In the case of Kepler-90g, a 
more useful constraint on the planet's density comes from demanding the moon 
orbit within the Hill sphere, which requires

%\begin{align}
%\frac{a_{SP}}{R_P} \frac{R_P}{R_{\star}} < \frac{a}{R_{\star}} \sqrt[\leftroot{-1}\uproot{2}%\scriptstyle 3]{\frac{M_P}{3 M_{\star}}},
%\end{align}
%
%which may be re-arranged to

\begin{align}
\Bigg(\frac{M_P}{M_{\star}}\Bigg) > 3 \Bigg( \frac{ (R_P/R_{\star}) (a_{SP}/R_P) }{ (a/R_{\star}) } \Bigg)^{3},
\end{align}

where the bracketed terms on the right-hand side are all observables from our
light curve model. Using our posterior distributions, this constraint imposes
$(M_P/M_{\star}) > (8.4_{-2.1}^{+3.1})\times 10^{-5}$. Dynamically
speaking, we estimate the seven-planet system is stable if Kepler-90g has a mass 
$(M_P/M_{\star})\lesssim2\times10^{-4}$, implying the Hill sphere can plausibly 
extend to encompass this moon signal.

Since $\rho_P = \rho_{\star} (M_P/M_{\star}) (R_P/R_{\star})^{-3}$, then our 
limit may be expressed as $\rho_P > 0.39_{-0.10}^{+0.14}$\,g\,cm$^{-3}$, where 
we have used $\rho_{\star} = (1.036\pm0.033)$\,g\,cm$^{-3}$, based on averaging 
the light curve derived stellar densities from the seven planets in the system.
Densities greater than this sub-Saturn mean density are certainly plausible for
an $8$\,$R_{\oplus}$ planet. Additionally, for a retrograde moon, 
\citet{domingos:2006} argue moons are stable up to nearly the edge of the Hill
sphere ($\sim93$\%).

\subsection{Conclusions}
\label{sub:photoconclusions}

Based on our light curve analysis, we draw the following conclusions. Firstly,
the light curve can be well-explained by Kepler-90g hosting a single
large exomoon. Although the formal significance is $5.2$\,$\sigma$, we find that
the significance is dominated by the single moon-like transit in Q6, which 
indicates the need for a careful inspection of this event. Secondly, we
find that Kepler-90g.01 can be within the Hill sphere of the Kepler-90g, if
Kepler-90g has a mean density in the range of 
$0.39\rightarrow0.93$\,g\,cm$^{-3}$ (upper limit based on seven-planet 
stability), which is certainly plausible given the radius of $8$\,$R_{\oplus}$.

\section{Centroid Analysis}
\label{sec:centroids}

\subsection{Conventional Centroids}
\label{sub:conventional}

We have now established that the case for an exomoon around Kepler-90g now rests
solely on the reality of the Q6 moon-like transit. The usual detection criteria
used by the HEK project \citep{hek:2013} are less useful for this system, since 
for example we cannot conduct double-likelihood tests by omitting some of the 
transits, given that there is just one moon-like signal. Low amplitude 
transit-like events are sometimes imparted into the \kepler\ time series due to 
various sources of false positives, ranging from astrophysical to instrumental 
\citep{tenenbaum:2014}.

We confirmed that this signal is coherent in both the PDC and Simple Aperture 
Photometry (SAP) data products, using both \cofiam\ and polynomial detrendings 
of each. The moon-like signal is therefore not an artifact of the PDC
pipeline, which can occur in rare instances \citep{christiansen:2013}. Using
the Pixel Response Function (PRF; see \citealt{bryson:2010}), precise
centroids of each \kepler\ target are made available as a standard data product.
These may be used to identify certain types of false positive scenarios, such
as a background eclipsing binary, and this has become a standard component of 
the Data Validation (DV) process by the \kepler\ team \citep{bryson:2013}. 
During a transit, a large centroid shift indicates that the true eclipsing 
object is offset from the total flux centroid.

\begin{figure}
\begin{center}
\includegraphics[width=9.5 cm]{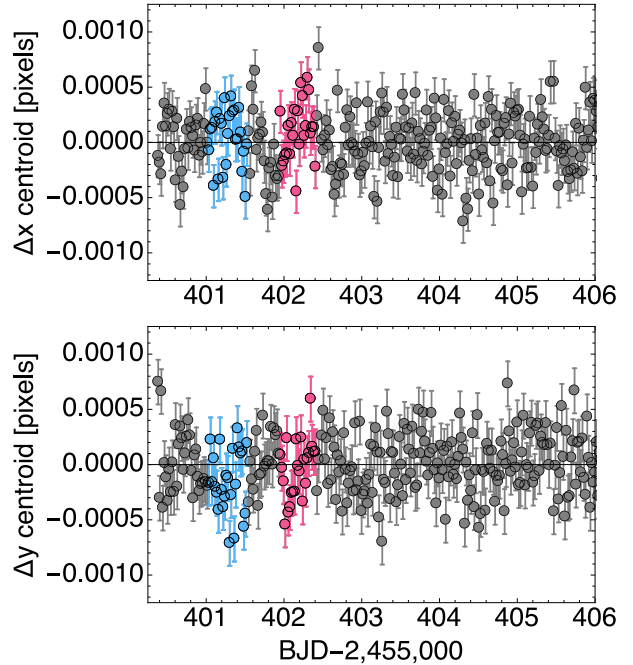}
\caption{\emph{
Time series of Kepler-90's centroids around the time of Kepler-90g's transit 
in Q6. Blue points are those occurring during the transit of Kepler-90g and
pink are those occurring during the transit of the putative exomoon 
Kepler-90g.01.
}} 
\label{fig:classiccentroids}
\end{center}
\end{figure}

In Fig.~\ref{fig:classiccentroids}, we show the time series of the $x$ (column)
and $y$ (row) centroids for Kepler-90 during Q6, after detrending with \cofiam. 
The highlighted areas depict the times when Kepler-90g and Kepler-90g.01 are 
transiting. Both signals appear reasonably consistent with the out-of-transit 
centroids at a level $<3$\,$\sigma$ (the typical threshold in DV reports). We 
find no compelling reason to reject the possible moon Kepler-90g.01 based on 
this information.

\subsection{``Transit Centroids''}
\label{sub:transitcentroid}

We wished to understand where on the CCD the transits occur and how localized
the transit signals were relative to the point spread function. In order to 
investigate this, we considered a new type of centroid we dub the ``transit 
centroid''. Whereas a conventional centroid calculates the mean pixel position 
weighted by the flux of each pixel, the transit centroid instead weights by the 
signal-to-noise ratio (SNR) of the transit depth on each pixel. Since the flux 
drops away from the flux centroid, causing greater photon noise, the SNR should 
be located at the same position as the flux centroid. If high SNR transits are 
found far away from the flux centroid, this is an indication that the transit is 
not localized at the expected flux centroid position. We define the transit 
centroid position as:

\begin{align}
x_{\mathrm{transit}} &= \frac{ \sum_{i=1}^N x_i (\delta_i/\sigma_i) }{ \sum_{i=1}^N (\delta_i/\sigma_i) } \\
y_{\mathrm{transit}} &= \frac{ \sum_{i=1}^N y_i (\delta_i/\sigma_i) }{ \sum_{i=1}^N (\delta_i/\sigma_i) },
\end{align}

where \{$x_i$,$y_i$\} is the Cartesian location of the center of the 
$i^{\mathrm{th}}$ pixel, $\delta_i$ is the transit depth recorded on that pixel 
and $\sigma_i$ is the associated uncertainty on that transit depth. For the
depth calculation, we simply compute the mean flux within the durations
found by our earlier photodynamical fits and use the standard error as the
uncertainty.

%%% Q6 SQUARE PLOTS
\begin{figure*}[ht]
\subfigure[Total Flux Centroid
\label{fig:totalcentroid_q6}]
{\epsfig{file=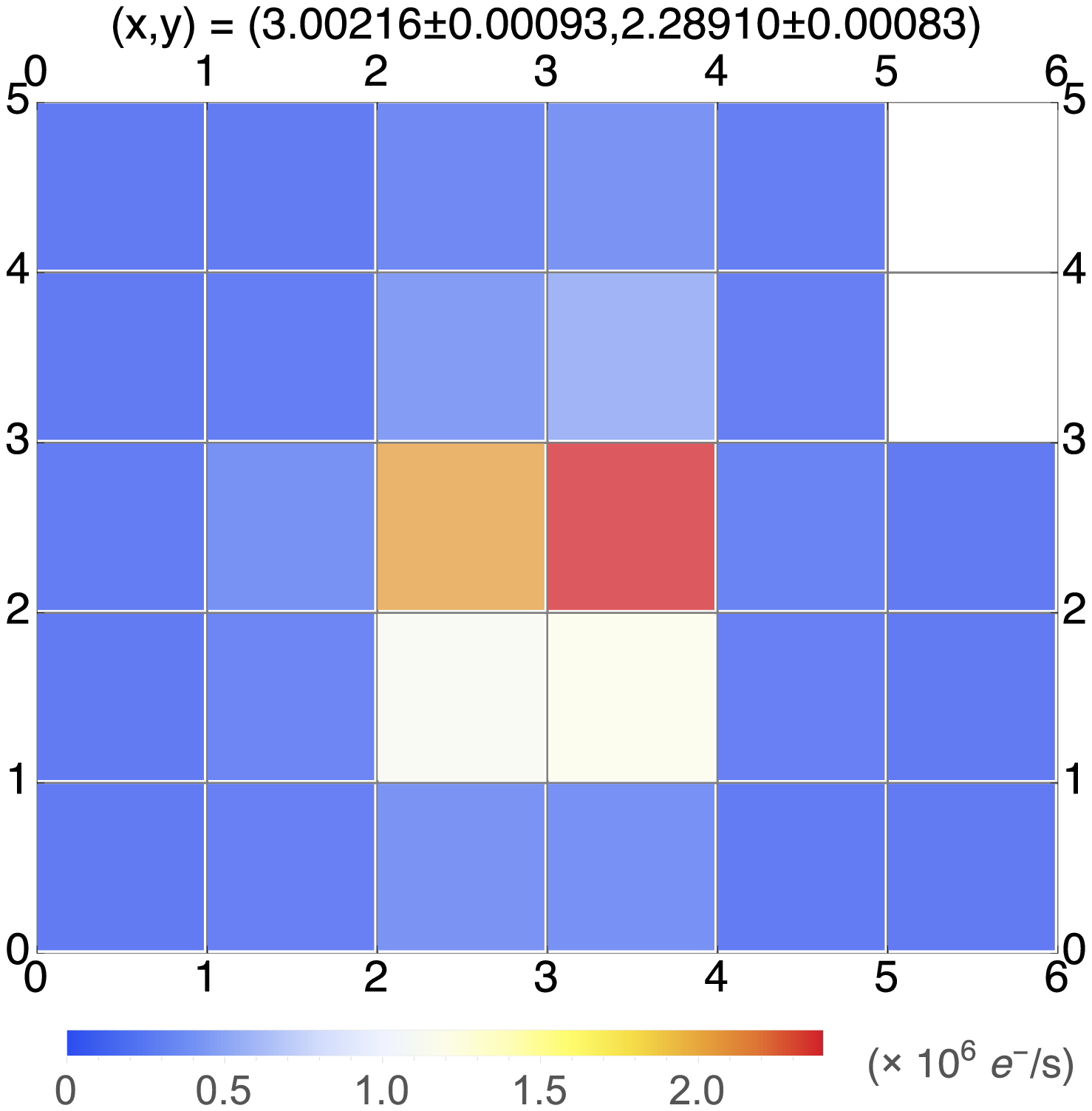,width=60mm}}
\subfigure[Kepler-90g Transit Centroid
\label{fig:plancentroid_q6}]
{\epsfig{file=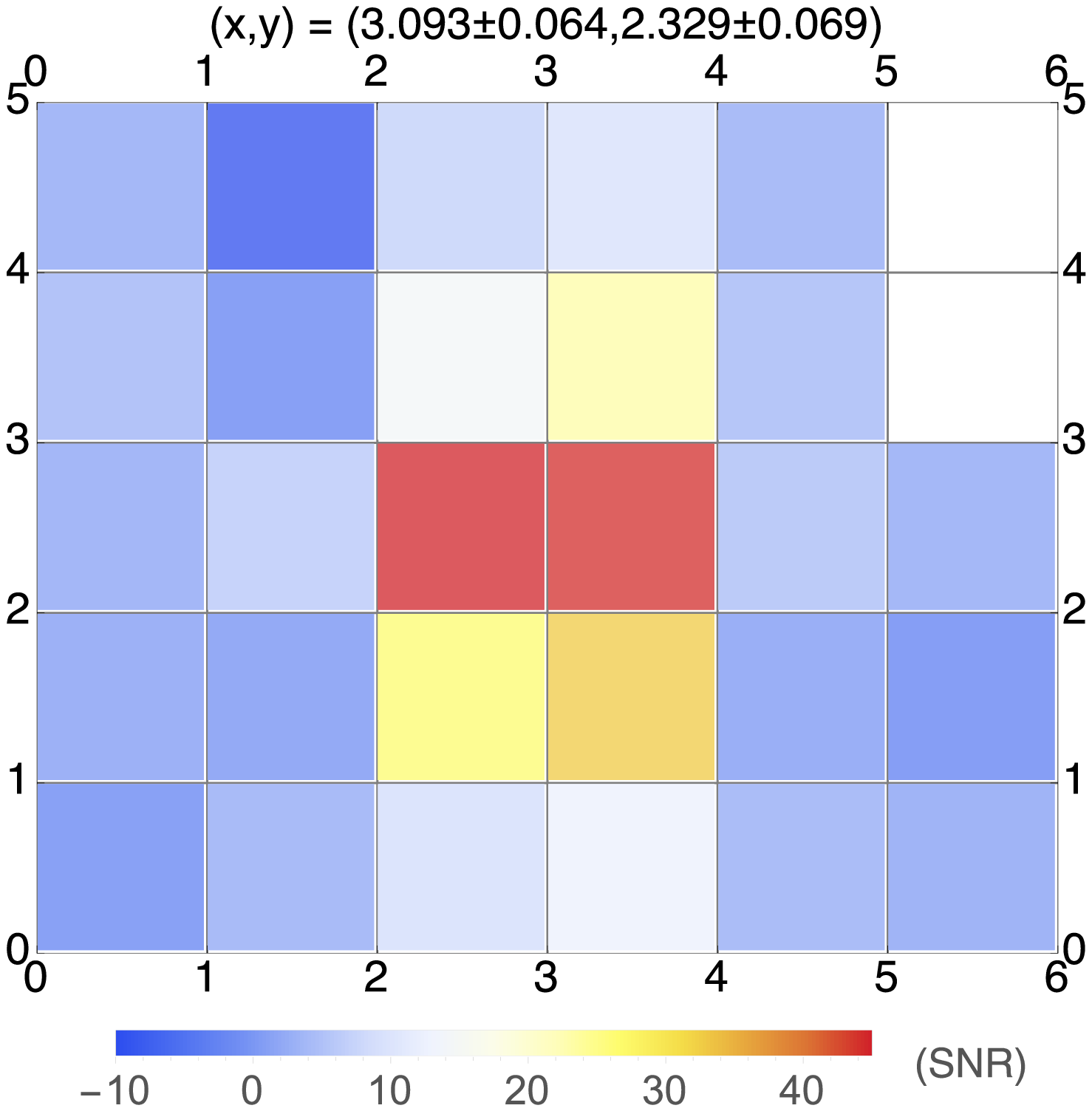,width=60mm}}
\subfigure[Kepler-90g.01 Transit Centroid
\label{fig:mooncentroid_q6}]
{\epsfig{file=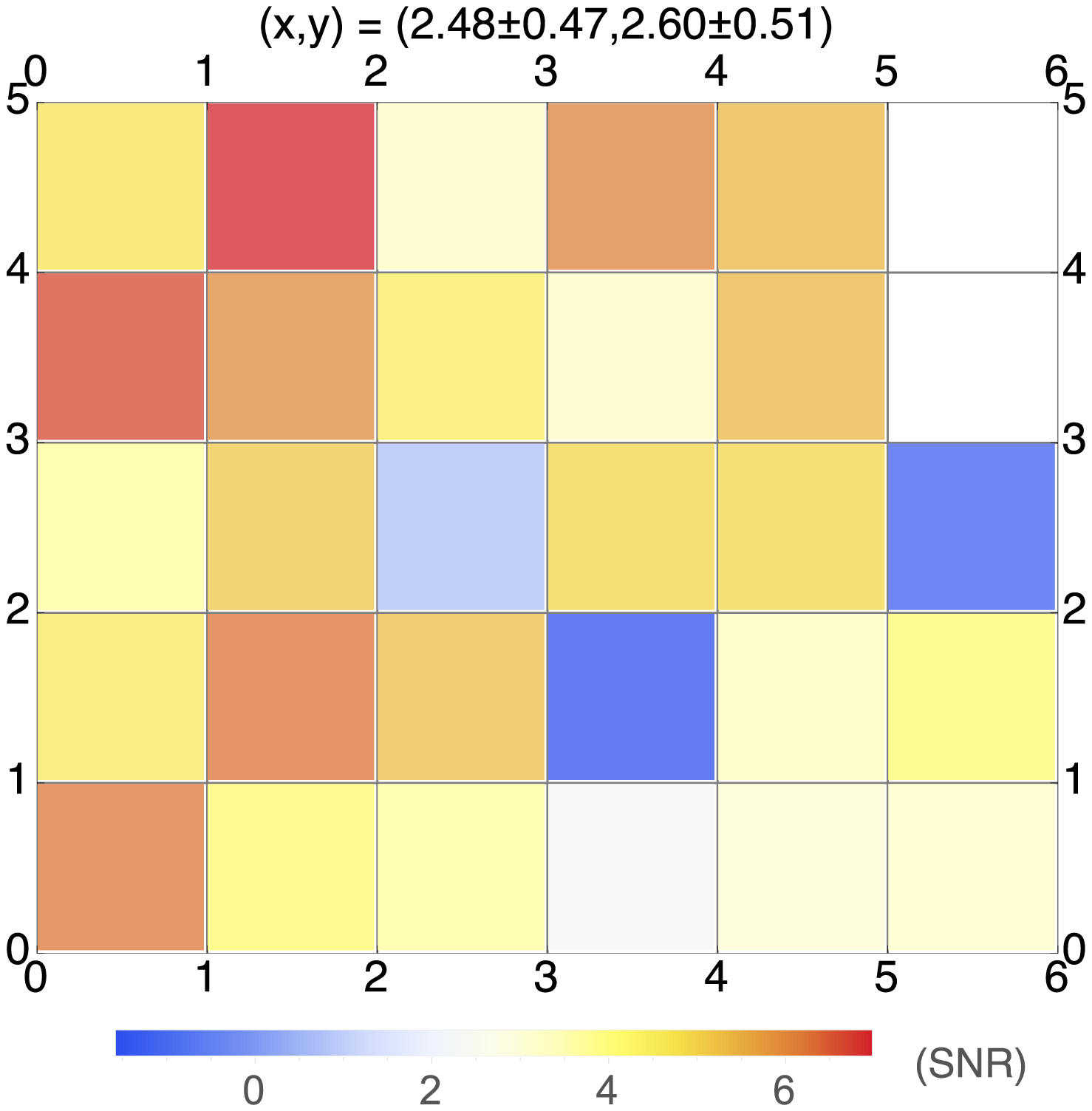,width=60mm}} \\
\caption{
Kepler pixel map of Kepler-90 during Q6. From blue to red, the color of each 
pixel is proportional to the signal-to-noise of the transit signal identified in 
this quarter, except for panel a) where the color scales with the flux. 
Kepler-90g.01 displays a highly delocalized transit centroid, making it doubtful 
as signal of astrophysical origin.
\label{fig:centroids_q6}}
\end{figure*}

%%% Q6f SQUARE PLOTS
\begin{figure*}
\subfigure[Total Flux Centroid
\label{fig:totalcentroid_q6f}]
{\epsfig{file=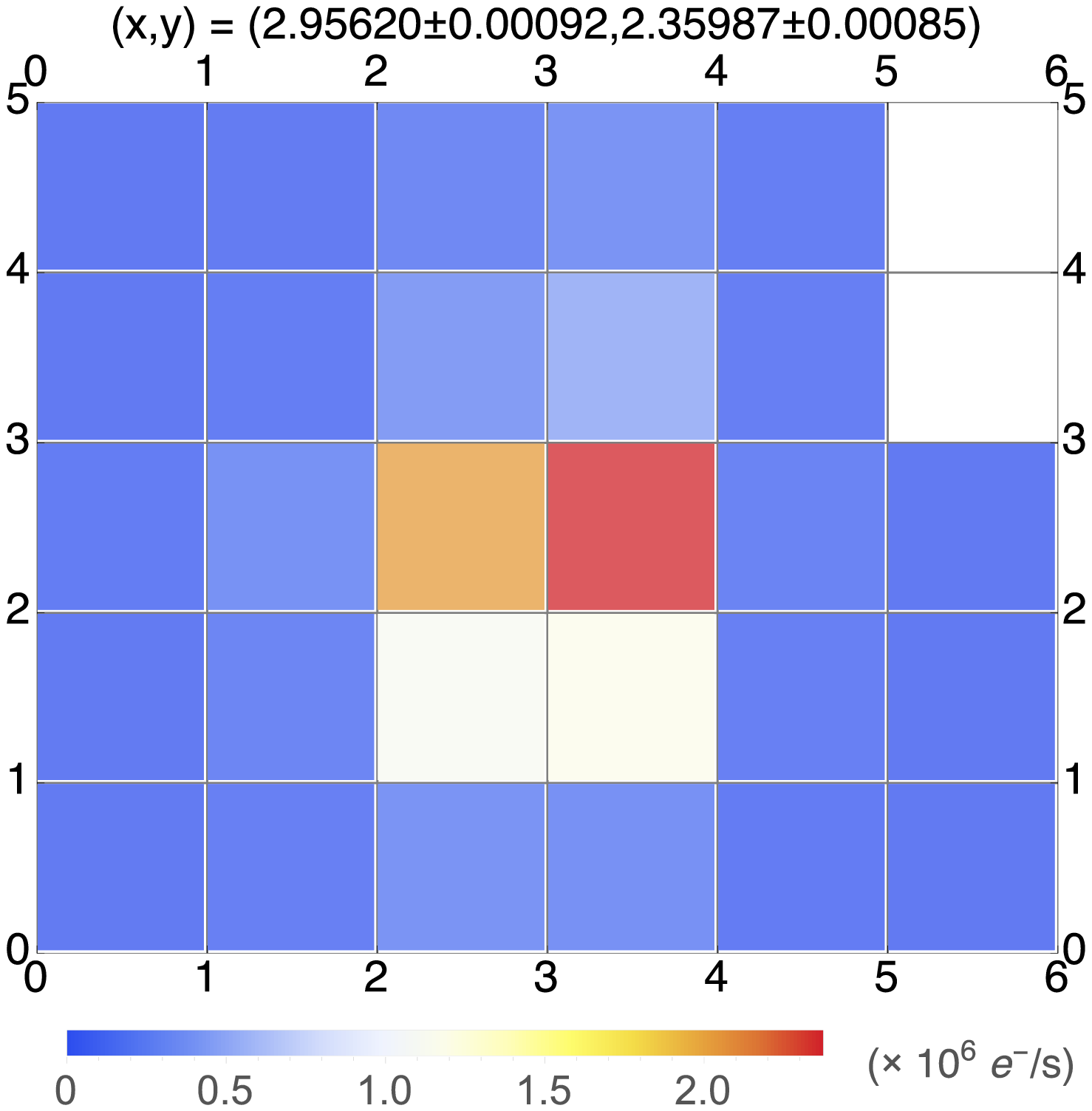,width=60mm}}
\subfigure[Kepler-90g Transit Centroid
\label{fig:plancentroid_q6f}]
{\epsfig{file=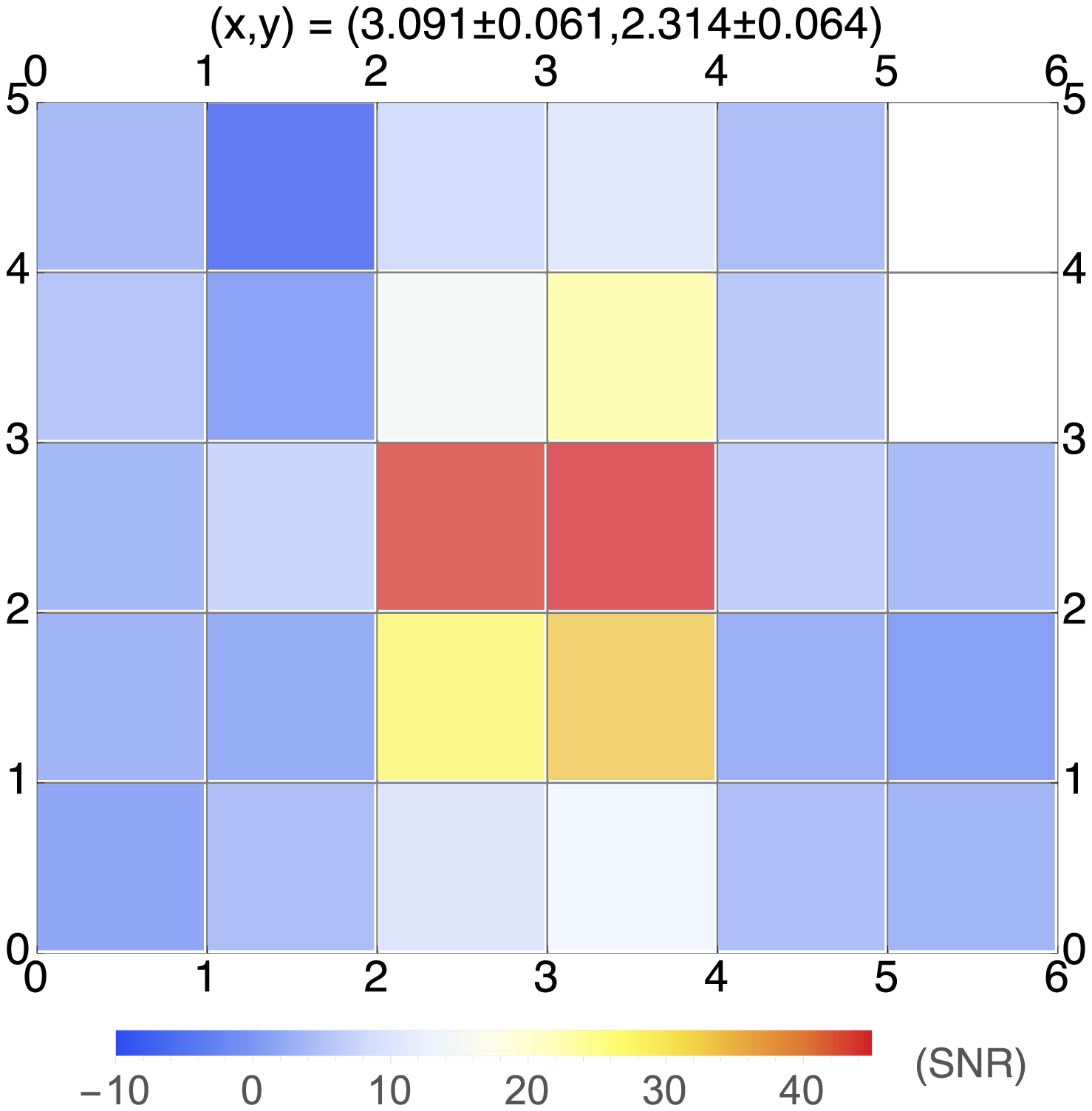,width=60mm}}
\subfigure[Kepler-90g.01 Transit Centroid
\label{fig:mooncentroid_q6f}]
{\epsfig{file=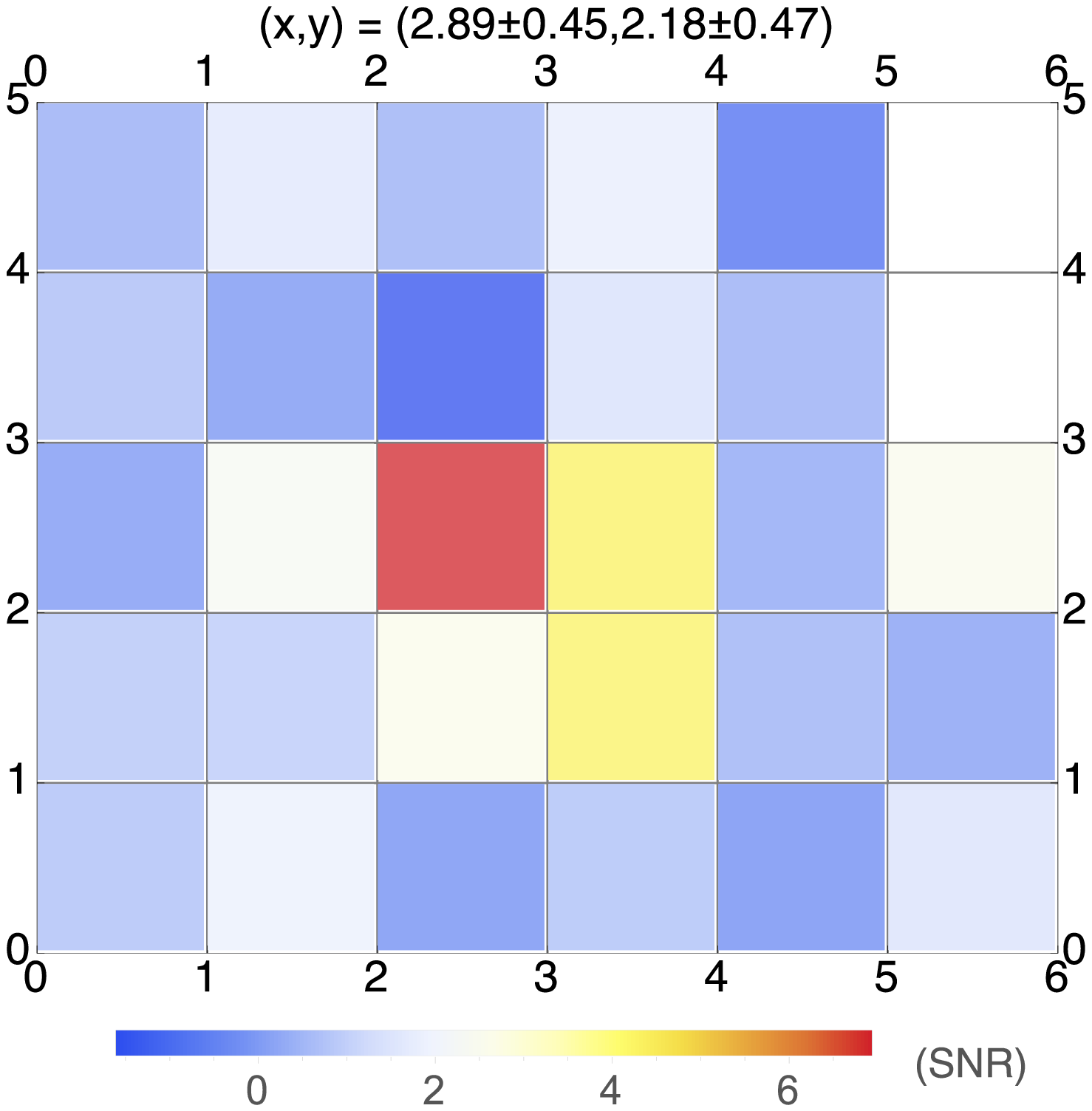,width=60mm}} \\
\caption{
Same as Fig.~\ref{fig:centroids_q6}, except we have replaced the moon candidate 
with an injected fake moon of equivalent transit depth. Unlike the real data,
the Kepler-90g.01 transit centroid is localized at the center of the postage
stamp, along with the Kepler-90g transit and the total flux.
\label{fig:centroids_q6f}}
\end{figure*}

The SNR of each pixel for the transits of Kepler-90g and Kepler-90g.01 during Q6 
are color coded on Fig~\ref{fig:plancentroid_q6} \& \ref{fig:mooncentroid_q6} 
respectively, which may be compared to the total flux centroid shown in 
Fig.~\ref{fig:totalcentroid_q6}. Centroid positions are provided at the top
of each panel in Fig~\ref{fig:centroids_q6}, with uncertainties estimated using
a bootstrap of each pixel's light curve residuals. Although the Kepler-90g.01 
transit centroid is not significantly offset ($<2$\,$\sigma$) from the
planetary transit, the SNR pattern for Kepler-90g.01 is quite distinct. 
Specifically, it is clear that the moon-like signal is not localized at the 
center of the postage stamp, unlike Kepler-90g, with even pixels on the edges 
showing high SNR levels.

We considered the possibility that the fact Kepler-90g.01 is such a low
amplitude transit is responsible for the transit centroid's delocalization,
i.e. the data are not good enough to pin down the location. To test this, we
created a fake Q6 time series by taking the pixel-level light curves of 
Kepler-90g's transit, reducing the depth to equal that of Kepler-90g.01, and
injecting it in place of the original Kepler-90g.01 transit for each pixel. 
Since the durations of the events are similar ($12.1$\,hours for g and 
$10.4$\,hours for g.01), this process should produce a realistic estimate of the 
transit centroid pattern produced if Kepler-90g.01 were a genuine exomoon. 
Recalculating the transit centroids (see Fig.~\ref{fig:centroids_q6f}), we find 
that of the fake moon to be highly localized, as visible in 
Fig.~\ref{fig:mooncentroid_q6f}, and the transit centroid is now within 
1\,$\sigma$ of both Kepler-90g and the total flux. This test indicates that the 
delocalized structure seen in Fig.~\ref{fig:mooncentroid_q6} is strong evidence 
that the Kepler-90g.01 transit is spurious.

On this basis, we identify Kepler-90g.01 as a false positive exomoon. The
underlying origin for the anomalous transit centroid is unclear, but may be due
to a cosmic ray hit causing Sudden Pixel Sensitivity Dropout (SPSD) at this 
time (see \citealt{christiansen:2013}). As a further check, we tried detrending 
the Q6 photometry using the Trend Filtering Algorithm (TFA) described in 
\citet{huang:2013}, which searches for common mode behaviors on the same CCD 
module. We find that the moon-like transit persists in the TFA light curve, 
indicating that the other stars on this module do not exhibit the event.
For the purposes of this short letter it is unimportant what the actual origin 
is; the key conclusion is that the signal is unlikely to be that of an exomoon.

\section{Discussion}
\label{sec:discussion}

In this work, we have demonstrated that the exomoon-like transit signal
identified near one of the transits of the 8\,$R_{\oplus}$ exoplanet Kepler-90g 
by \citet{cabrera:2014} is likely a false positive. This determination is far 
from obvious, having required us to introduce a new tool to examine \kepler\ 
pixel-level data.

\citet{cabrera:2014} originally suggested the signal may be spurious based on 
the large temporal separation between the transit of Kepler-90g and the 
moon-like transit. However, in this work, we find that the separation 
corresponds to a moon within the Hill sphere, provided Kepler-90g has a mean 
density exceeding 0.39\,g\,cm$^{-3}$. Further, a retrograde exomoon would be 
stable up to nearly the edge of the Hill sphere, following 
\citet{domingos:2006}. Based on dynamical stability simulations, the 
seven-planet system in which Kepler-90g resides is stable provided g has a 
density below 0.93\,g\,cm$^{-3}$. Finally, we find that the fact that only one 
of the six observed transits of Kepler-90g shows a high signal-to-noise exomoon 
transit is easily explained with a photodyanmical model. Based on a purely
photometric analysis then, one would conclude that this is a plausible exomoon
candidate.

Analysis of the PRF centroids shows no significant deviations during the times
of Kepler-90g's transit, nor that of the moon-like signal. Inspecting the
pixel-level data however, we suspected that the moon-like transits were 
occurring on nearly all pixels rather than being centered on the flux centroid. 
To test this, we considered a new test, dubbed the ``transit centroid'', which 
computes a centroid position weighted by the signal-to-noise of the transit, as 
seen by each pixel. This test confirms that the moon-like transit is not 
localized at the expected position, a result which we demonstrate is not due to 
the low amplitude nature of the moon-like signal either. We conclude that the 
moon-like transit is likely an instrumental artifact, perhaps due to a cosmic 
ray causing a Sudden Pixel Sensitivity Dropout \citep{christiansen:2013}.

This work establishes that the moon-like signal near Kepler-90g is likely 
spurious and does not consititute compelling evidence for a transiting exomoon.
The fact that the candidate passes all of the conventional tests, such as 
photodynamical modeling, imposing dynamical stability and centroid analysis, 
highlights the extreme difficulty facing exomoon hunters, due to the 
non-periodic and low-amplitude nature of the signals being sought. Nevertheless, 
the sensitivity to sub-Earth mass exomoons using \kepler\ is firmly established 
and despite the mine-field of false positives, a confirmed detection is surely 
inevitable.

% #####################################################################
%% Acknowledgements
\acknowledgements
\section*{Acknowledgements}

% Dodds thanks
This work made use of the Michael Dodds Computing Facility.
% Personal funding
DMK is funded by the CfA Menzel Fellowship. GB acknowledges partial 
support from NSF grant AST-1108686 and NASA grant NNX12AH91H.
%
%% EOF Acknowledgements

% #####################################################################
%% Bibliography


\begin{thebibliography}{99}
\bibitem[\protect\citeauthoryear{Barclay et al.}{2013}]{barclay:2013} Barclay,
T., Rowe, J. F., Lissauer, J. J., et al., 2013, Nature, 494, 452
\bibitem[\protect\citeauthoryear{Bennett et al.}{2014}]{bennett:2014} Bennett,
D. P., Batista, V., Bond, I. A., et al., 2014, ApJ, 785, 155
\bibitem[\protect\citeauthoryear{Bryson et al.}{2010}]{bryson:2010}
Bryson, S. T., Tenenbaum, P., Jenkins, J. M., et al., 2010, ApJ, 713, L97
\bibitem[\protect\citeauthoryear{Bryson et al.}{2013}]{bryson:2013}
Bryson, S. T., Jenkins, J. M., Gilliland, R. L., et al., 2013, PASP, 125, 889
\bibitem[\protect\citeauthoryear{Cabrera et al.}{2014}]{cabrera:2014} Cabrera,
J., Csizmadia, Sz., Lehmann, H. et al., 2014, ApJ, 781, 18
\bibitem[\protect\citeauthoryear{Christiansen et al.}{2013}]{christiansen:2013} 	
Christiansen, J. L., Clarke, B. D., Burke, C. J., et al., 2013, ApJS, 207, 35
\bibitem[\protect\citeauthoryear{Domingos et al.}{2006}]{domingos:2006}
Domingos, R. C., Winter, O. C. \& Yokoyama, T., 2006, MNRAS, 373, 1227
\bibitem[\protect\citeauthoryear{Feroz et al.}{2008}]{feroz:2008} 
Feroz, F. \& Hobson, M. P., 2008, MNRAS, 384, 449
\bibitem[\protect\citeauthoryear{Feroz et al.}{2009}]{feroz:2009} 
Feroz, F., Hobson, M. P. \& Bridges, M., 2009, MNRAS, 398, 1601
\bibitem[\protect\citeauthoryear{Huang et al.}{2013}]{huang:2013} 
Huang, X., Bakos, G. A. \& Hartman, J. D., 2013, MNRAS, 429, 2001
\bibitem[\protect\citeauthoryear{Kipping}{2009}]{kipping:2009a} Kipping, 
D. M., 2009, MNRAS, 392, 181
\bibitem[\protect\citeauthoryear{Kipping}{2010}]{weighing:2010} Kipping, 
D. M., 2010, MNRAS, 409, L119
\bibitem[\protect\citeauthoryear{Kipping}{2011}]{luna:2011} Kipping, 
D. M., 2011, MNRAS, 416, 689
\bibitem[\protect\citeauthoryear{Kipping et al.}{2013a}]{hek:2013} Kipping, 
D. M., Hartman, J., Buchhave, L. A., et al., 2013a, ApJ, 770, 101
\bibitem[\protect\citeauthoryear{Kipping et al.}{2013b}]{kepler22:2013} Kipping, 
D. M., Forgan, D., Hartman, J., et al., 2013b, ApJ, 777, 134
\bibitem[\protect\citeauthoryear{Kipping et al.}{2014}]{hek:2014} Kipping, 
Nesvorn\'y, D., Buchhave, L. A., et al., 2014, ApJ, 784, 28
\bibitem[\protect\citeauthoryear{Schmitt et al.}{2014}]{schmitt:2014} Schmitt,
J. R., Wang, J., Fischer, D. A., et al., 2014, AJ, 148, 28
\bibitem[\protect\citeauthoryear{Schneider et al.}{2011}]{schneider:2011} 
Schneider, J., Dedieu, C., Le Sidaner, P., Savalle, R. \& Zolotukhin, I., 2011,
A\&A, 532, 79
\bibitem[\protect\citeauthoryear{Smith et al.}{2012}]{smith:2012} 
Smith, J. C., Stumpe, M. C., Van Cleve, J. E., et al., 2012, PASP, 124, 1000
\bibitem[\protect\citeauthoryear{Tenenbaum et al.}{2014}]{tenenbaum:2014} 
Tenenbaum, P., Jenkins, J. M., Seader, S., et al., 2014, ApJS, 211, 6
\end{thebibliography}
\end{document}